# The Cool Giant HD 77361 - A Super Li-Rich Star

L. S. Lyubimkov[1], B. M. Kaminsky[2], V. G. Metlov[3],
Ya. V. Pavlenko[2], D. B. Poklad[1], and T. M. Rachkovskaya[1]

[1]*Crimean Astrophysical Observatory, pos. Nauchny, Crimea, 298409 Russia*
[2]*Main Astronomical Observatory, National Academy of Sciences of Ukraine, Kiev, Ukraine*
[3]*Sternberg Astronomical Institute, Moscow State University, Universitetskii pr. 13, Moscow, 119991 Russia*

**Abstract**—Super Li-rich stars form a very small and enigmatic group whose existence cannot be explained in terms of the standard stellar evolution theory. The goal of our study is to check the reality of this group of cool giants based on an independent technique. We have carried out such a check using the K giant HD 77361 (HR 3597), which has previously been assigned to this rare type, as an example. We have redetermined the effective temperature $T_{eff}$ and surface gravity $\log g$ for this star. We have applied two different methods, photometric and spectroscopic, to estimate $T_{eff}$ (the accuracy of the Li-abundance determination depends significantly on this parameter). The value of $\log g$ has been found from the highly accurate parallax of this nearby star. To apply the photometric method of determining $T_{eff}$, we have performed $UBV$ observations of the star, which yielded $V = 6.18 \pm 0.03$, $B - V = 1.13 \pm 0.01$, and $U - B = 1.18 \pm 0.05$. The following parameters of the star have been found: effective temperature $T_{eff} = 4370 \pm 100$ K, surface gravity $\log g = 2.30 \pm 0.10$, iron abundance $\log \varepsilon(Fe) = 7.49 \pm 0.14$, microturbulence $V_t = 1.1 \pm 0.2$ km s$^{-1}$, rotational velocity $V\sin i = 4.5$ km s$^{-1}$, and mass $M = 1.3 \pm 0.2$ M$_\odot$. The lithium abundance has been determined from a non-LTE analysis of three Li I lines: the resonance line at 6707.8 Å and the subordinate lines at 6103.6 and 8126.4 Å (the latter in a blend with a CN molecular line). We have found a high lithium abundance, $\log \varepsilon(Li) = 3.75 \pm 0.11$, which exceeds considerably the initial abundance $\log \varepsilon(Li) = 3.2 \pm 0.1$ for young stars in the solar neighborhood. Thus, we have confirmed that the K giant HD 77361 actually belongs to the type of super Li-rich stars. It is noted that a high lithium abundance in such cool giants is inconsistent with predictions of the standard stellar evolution theory and may suggest a recent synthesis of lithium in these stars.

## INTRODUCTION

Lithium is an extremely sensitive indicator of stellar evolution. Therefore, this element invariably attracts the particular attention of researchers. The Li I lines from which the lithium abundance is determined are seen only in the spectra of sufficiently cool F-M stars (only the strongest, resonance Li I 6707.8 Å line is often seen in the spectrum). Therefore, the Li abundances derived from observations mainly refer precisely to such stars.

The lithium abundance in the stellar atmosphere can be reduced significantly relative to its initial value of $\log \varepsilon(Li) = 3.2 \pm 0.1$, typical for young stars in the solar neighborhood, already at the first and longest evolutionary phase, when hydrogen burns in the stellar core (this is the main-sequence (MS) phase). This follows from the observations of F, G, and K dwarfs, including the Sun. Such a reduction is explained by mixing in the MS phase, which causes the material from the stellar atmosphere to fall into deeper and hotter layers, where the lithium atoms are destroyed (lithium depletion begins at a temperature $\sim 2.5 \times 10^6$ K). A more dramatic reduction in the lithium abundance is observed in F, G, and K giants and supergiants, which is explained by deep convective mixing in this evolutionary phase. This phase is of particular interest to us. In what follows, the lithium abundance $\log \varepsilon(Li)$ is given in an ordinary logarithmic scale, where $\log \varepsilon(H) = 12.00$ is adopted for hydrogen.

The overwhelming majority (up to 90%) of FGK giants and supergiants do not show the Li I 6707.8 Å line in their spectra at all, i.e., there is little lithium in their atmospheres or it has been completely depleted as a result of mixing (see, e.g., Liu et al. 2014). An explanation for this phenomenon was given already in the first studies aimed at determining the Li abundance in stars of the type under consideration (see, e.g., Luck 1977; Lambert et al. 1980). As has already been noted, the cause of the observed Li deficiency is deep convective mixing in this evolutionary phase. Interestingly, the same cause leads to a simultaneous reduction in the $^{12}C/^{13}C$ ratio. Comparison with predictions of the theory in those years was limited by the fact that only the stellar models without rotation could be calculated. A detailed comparison with the present-day theoretical models computed both with and without rotation can be found in Lyubimkov et al. (2012, hereafter LLKPPR'12). Note that the Li I 6707.8 Å line profile was analyzed in this paper by abandoning the LTE (local thermodynamic equilibrium) condition, which, in general, is important for a reliable determination of the Li abundance. The non-LTE corrections to the lithium abundance $\log \varepsilon(Li)$ are known to depend on specific parameters of the star, including its effective temperature $T_{eff}$, surface gravity $\log g$, metallicity index [Fe/H], microturbulence $V_t$, and the value of $\log \varepsilon(Li)$ itself (Lind et al. 2009).

In LLKPPR'12, the Li abundance was found for 55 Galactic F and G supergiants and luminous giants (stars of luminosity classes I and II). Comparison with theoretical models showed the atmospheric Li abundance to depend strongly on the initial rotational velocity of the star $V_0$ and its mass $M$. In particular, according to LLKPPR'12, the models without rotation ($V_0 = 0$ km s$^{-1}$), which, on average, give log ε(Li) = 1.4, are well suited for stars with masses $M < 6\ M_\odot$ that have lithium abundances log ε(Li) = 1.1 - 1.8. However, even at a low initial rotational velocity Vo ~ 50 km s$^{-1}$, the Li abundance becomes undetectably small (log ε(Li) < 1) after mixing in the FGK-giant phase.

The small group of Li-rich stars occupies a special place among the FGK giants and supergiants; the existence of at least some of them is difficult to explain in terms of the standard stellar evolution theory. The objects with abundances log ε(Li) > 2.0 belong to such stars. These stars are encountered rarely; their fraction among the cool giants is ~1% (Uttenthaler et al. 2012; Liu et al. 2014). The fact that there are few such stars may suggest a short duration of this evolutionary phase. According to LLKPPR'12, all of the known Li-rich giants and supergiants have an upper limit on the mass: $M < 6\ M_\odot$. It should be noted that an enhanced lithium abundance log ε(Li) ~ 2-3 is observed, in particular, in cool stars with an enhanced chromospheric activity (see, e.g., Takeda et al. 2010); however, the star HD 77361 being investigated below shows no signatures of such an activity.

Remarkably, several cool giants and supergiants with an extremely high lithium abundance, log ε(Li) ~ 4, have been detected among the Li-rich stars. For example, Kumar and Reddy (2009) collected data for five K giants with log ε(Li) = 3.7-4.2; they assigned such stars to the type of super Li-rich ones. Since the lithium abundance in such giants exceeds appreciably its initial value of log ε(Li) = 3.2 ± 0.1, it has to be assumed that recently synthesized lithium is observed in their atmospheres; however, as Kumar and Reddy (2009) pointed out, this assumption runs into serious difficulties.

The goal of this study is to confirm or question the reality of the small and enigmatic group of super Li-rich stars, whose existence cannot be explained in terms of the standard stellar evolution theory. Since the super Li-rich stars are very few in number, the question about the accuracy of the log ε(Li) estimates obtained for them inevitably arises. Maybe, the Li abundance in these rare cases was erroneously overestimated? For example, the discrepancy in the Li abundances found for the same stars by different authors is known to reach 0.4-0.5 dex (see, e.g., Fig. 3 in LLKPPR'12). If such an error in the Li abundance is admitted for the mentioned five K giants from Kumar and Reddy (2009), then the Li overabundances found for them will be reduced significantly. To make sure that super Li-rich stars actually exist, it is first necessary to check how accurate the adopted parameters $T_{eff}$ and log $g$ for the stars assigned to this rare type are. In particular, the accuracy of the adopted effective temperature $T_{eff}$ is important, because the intensity of the resonance Li I 6708 Å line, from which the Li abundance in cool stars is usually estimated, depends significantly on it.

We carried out such a check using one specific star assigned by Kumar and Reddy (2009) to this type as an example. This is the K giant HD 77361, the brightest super Li-rich star among the five stars considered by these authors. Kumar and Reddy found the effective temperature $T_{eff}$ = 4580 K, surface gravity log $g$ = 2.5, microturbulence $V_t$ = 1.4 km s$^{-1}$, and lithium abundance log ε(Li) = 3.82 for HD 77361. Apart from the resonance Li I 6707.8 Å line, they also used the subordinate Li I 6103.6 Å line to estimate log ε(Li). The non-LTE corrections to the derived abundances log ε(Li) were estimated using published data.

We performed a comprehensive study of the giant HD 77361 anew based on our own technique, including the determination of its fundamental parameters, the investigation of its metallicity, and a non-LTE analysis of the lithium abundance. Our technique differs fundamentally from that applied by Kumar and Reddy. In our opinion, it is this approach that allows the status of the star HD 77361 to be checked independently.

Note that the star HD 77361 (HR 3597) was classified as a K1 III giant. The Hipparcos satellite obtained the parallax π = 9.25 ± 0.43 mas for it (van Leeuwen 2007), implying $d = 108 ± 5$ pc. The star lies at a comparatively short distance from the Sun, with its apparent magnitude being = 6.2.

In our study, we use the high-resolution spectrum of HD 77361 that was taken by D.L. Lambert with the 2.7-m telescope at the MacDonald Observatory of the University of Texas (USA) and that was kindly placed by him at our disposal. The spectroscopic data used in LLKPPR'12, where a more detailed description of the observations can be found (see also Lyubimkov et al. 2010), was also obtained with the same telescope. Significantly, the same spectrum of HD 77361 was used in the cited paper by Kumar and Reddy (2009).

THE TECHNIQUE FOR DETERMINING THE FUNDAMENTAL PARAMETERS $T_{eff}$ AND log $g$

The effective temperature $T_{eff}$ and surface gravity log $g$ are among the fundamental (or basic) stellar parameters. They are known to be closely related to two other fundamental quantities, the mass $M$ and luminosity L. A study of the chemical composition for any star begins with the determination of $T_{eff}$ and log $g$, and the accuracy of the elemental abundances being determined depends on their accuracy. In particular, the effective temperature $T_{eff}$, to which the Li I lines are particularly sensitive (they depend weakly on log $g$), plays an important role in analyzing the lithium abundance. Therefore, we should briefly consider how $T_{eff}$ was determined in Kumar and Reddy (2009) and how this is done in our work.

Kumar and Reddy found $T_{eff}$ from Fe I lines (a total of 40 lines); this required the fulfilment of the following condition: the lines with different excitation potentials $E_l$ must give, on average, the same Fe abundance, i.e., there must be no trend in the Fe abundance with increasing $E_l$. It is important that, in this case, the possible departures from LTE were disregarded in the calculations of Fe I lines.

What non-LTE effects in Fe I lines might be expected for the K giant HD 77361? There was no answer to this question when we began our study of this star. Nor does it exist now, because there are no detailed non-LTE calculations of Fe I lines for K giants and supergiants. Such calculations were performed previously for K dwarfs; they showed the non-LTE correction to the iron abundance log ε(Fe) for such stars to be ~0.1 dex (see below). However, it is well known from numerous non-LTE calculations for other elements that the non-LTE corrections can increase when passing from dwarfs to giants and supergiants (i.e., as log $g$ decreases). Interestingly, the non-LTE corrections in Fe I lines for G and K dwarfs can depend on the excitation potential $E_l$, as was shown by Shchukina and Trujillo Bueno (2001) using the Sun as an example. Meanwhile, as has been noted above, the dependence of log ε(Fe) on $E_l$ was used by Kumar and Reddy as the basis for their technique of determining $T_{eff}$. All of what has been said above suggests that the neglect of non-LTE effects in Kumar and Reddy (2009) could lead to a significant error in the $T_{eff}$ estimate.

We obtained another argument against the use of Fe I lines to determine $T_{eff}$ for the cool giant HD 77361 when analyzing the iron abundance for this star (see below). Our calculations showed the Fe I lines to be barely sensitive to $T_{eff}$ variations precisely in the temperature range $T_{eff}$ = 4300-4500 K, within which the star being investigated falls. In other words, the Fe I lines cannot serve as a reliable indicator of $T_{eff}$ in this $T_{eff}$ range.

It should be noted that Kumar and Reddy bore out their spectroscopic estimate of $T_{eff}$ obtained from Fe I lines by their photometric estimate based on the $B - V$ and $V - K$ color indices. We will discuss these estimates below, when we will obtain the final value of $T_{eff}$ based on our technique.

As has already been noted in the Introduction, to confirm the status of the star HD 77361 with greater reliability, we applied a fundamentally different technique to determine $T_{eff}$. For this purpose, we used two independent methods: the recently developed photometric method of Lyubimkov and Poklad(2014) and the spectroscopic method of Kovtyukh (2007). The former was proposed to determine the effective temperature $T_{eff}$ for G and K giants and supergiants. It is based on the use of two observed photometric indices free from the influence of interstellar extinction: $Q = (U - B) - 0.72(B - V)$ in the photometric $UBV$ system and $[c_1] = c_1 - 0.20(b - y)$ in the $uvby$ system. This method allows one to find $T_{eff}$ from the $Q$ index in the temperature range 3800 K < $T_{eff}$ < 5100 K and from the $[c_1]$ index in the range 4900 K < $T_{eff}$ < 5500 K. Clearly, $T_{eff}$ can be determined for the cool giant HD 77361 by this method only from the $Q$ index. It should be noted that this method gives different dependences of $T_{eff}$ on $Q$ for stars with a normal (solar) metallicity, [Fe/H] = 0, and stars with a moderately reduced metallicity, [Fe/H] = -0.5.

As an example characterizing the accuracy of this method, Lyubimkov and Poklad (2014) determined $T_{eff}$ from the $Q$ index for Arcturus, a very close and thoroughly studied early K giant with metallicity [Fe/H] = -0.5. The value of $T_{eff}$ = 4262 ± 20 K found for it differed only by 24 K from $T_{eff}$ = 4286 ± 30 K derived by Ramirez and Allende Prieto (2011) from the spectral energy distribution of Arcturus in a wide wavelength range, from 0.44 to 10 $\mu$m. This example gives hope that this method also yields a reliable estimate of $T_{eff}$ for another, fairly close early K giant, HD 77361.

To determine the effective temperature of HD 77361, we also applied the other (spectroscopic) method (Kovtyukh 2007). It is based on an analysis of the ratio of the central intensities for pairs of lines with significantly differing excitation potentials. As was confirmed by Lyubimkov et al. (2010), this method gives a high accuracy of the $T_{eff}$ estimate for G and K supergiants and giants. The results of our $T_{eff}$ determination by both methods are considered below.

We found the second fundamental parameter, the star's surface gravity log $g$, from its trigonometric parallax $\pi$ by the method described in Lyubimkov et al. (2009, 2010). Owing to the high accuracy of the $\pi$ values obtained with the Hipparcos satellite (van Leeuwen 2007), this method of determining log $g$ may be considered to be the most accurate to date for relatively close stars.

*UBV* PHOTOMETRY FOR THE STAR
HD 77361

The $U - B$ and $B - V$ color indices should be known to determine $T_{eff}$ by the method of Lyubimkov and Poklad (2014) specifically from $Q = (U - B) - 0.72(B - V)$. In the catalogue by J. Mermilliod and M. Mermilliod (1994) and the SlMBAD database (http://simbad.u-strasbg.fr/simbad/ sim-fid), we found a reference only to one paper where $B - V$ = 1.13 is given for the star HD 77361 (Corben 1966); there is no $U - B$ for HD 77361 in the literature. Therefore, the first task to be accomplished was to carry out *UBV* observations of the star being investigated. Such observations were performed by one of us (V.G. Metlov) at the Crimean Station of the Sternberg Astronomical Institute (SAI), the Moscow State University, located in the neighborhood of the Crimean Astrophysical Observatory (CrAO). It should be noted that HD 77361 is a southern-sky star (its coordinates are $\alpha = 9^h01^m 11^s$ and $\delta = -26°39'$). Therefore, its observations at the CrAO latitude (44°44') were difficult to make because of its low position above the horizon.

Our photometric observations of HD 77361 were carried out with a 60-cm Zeiss telescope at the Crimean Station of the SAI over five nights, two nights in December 2013 and three nights in February 2014. A photon-counting *UBV* photometer (Lyuty 1971) was used for our measurements. There are no sufficiently bright stars with reliably measured *UBV* magnitudes that could serve as comparison stars in the immediate neighborhood of HD 77361, while its zenith distance is rather large even near the meridian. Therefore, we chose the observing technique that had long been used at the Crimean Station of the SAI to produce the catalogs of bright (4-8$^m$) reference stars for some astronomical space projects (Metlov 2002).

For the observations of HD 77361, we used only completely clear nights, when there were no clouds or inhomogeneous haze on the entire sky, down to the horizon, and the atmospheric transparency was stable. Such an approach was associated not only with the low position of the observed star above the horizon but also with the application of comparison stars and "extinction stars" from the highly accurate system of photometric standards produced at the Alma-Ata station of the SAI in 1970-1980 (Khaliullin et al. 1985). This system contains 72 standards on the entire sky, and the angular separations between them are sufficiently large.

On each observing night, we determined the atmospheric extinction coefficients at least two times by measuring two pairs of stars from the above system with different color indices *C*. The values of *C* in each pair are approximately identical, with one of the stars being near the zenith and the other being at a zenith distance of about 60°. The extinction coefficients in *U*, B, and *V* were found in the form $\alpha = \alpha_0 + \gamma C$ ($\alpha_0$ and $\gamma$ were determined). The results obtained were interpolated to the times of observations of the star being studied on a given night; thus, each magnitude estimate was obtained using its extinction coefficients. The above system of standards contains the *R, V*, B, and *W* magnitudes. We used *V* and *B* and took *U* from the BS catalog. The star HD 69830 was used as a standard for HD 77361.

We obtained 19 magnitude estimates (i.e., 19 independent pointings at the star) over five observing nights. All nights were distinguished by a good transparency; on average, $\alpha_V = 0.20$, $\alpha_B = 0.38$, and $\alpha_u = 0.80$ at *C* = 0. The accuracy of a single measurement in different filters under the above conditions was 0.010-0.015 in *V* and *B* and 0.03-0.04 in *U*. It should be noted that in analogous observations at zenith distances of less than 60°, the measurement accuracy turns out to be approximately a factor of 3 higher. In our case, however, the accuracy achieved for HD 77361 turned out to be quite acceptable for determining $T_{eff}$ (see below).

Table 1. Results of our *UBV* observations for HD 77361

| Date JD 2456000+ | V | B-V | U - B |
|---|---|---|---|
| 645.5337 | 6.172 | 1.129 | 1.110 |
| 650.5560 | 6.148 | 1.125 | 1.224 |
| 691.4407 | 6.174 | 1.135 | 1.204 |
| 693.4295 | 6.210 | 1.135 | 1.227 |
| 694.4192 | 6.210 | 1.113 | 1.156 |

Table 2. Determination of the parameters for HD 77361. To estimate $T_{eff}$, we used two different methods, photometric (Lyubimkov and Poklad 2014) and spectroscopic (Kovtyukh 2007)

| Method of determining $T_{eff}$ | $T_{eff}$ | log *g* | $M/M_\odot$ |
|---|---|---|---|
| Photometric at [Fe/H] = 0.0 | 4430 ± 90 | 2.39 ± 0.13 | 1.48 ± 0.21 |
| Photometric at [Fe/H] = -0.5 | 4270 ± 50 | 2.16 ± 0.07 | 1.14 ± 0.12 |
| Spectroscopic | 4310 ± 60 | 2.22 ± 0.08 | 1.22 ± 0.14 |

The values of *V*, *U - B*, and *B - V* averaged for each observing night are given in Table 1. Hence, on average, for HD 77361 we obtain

$$V = 6.18 \pm 0.03, \quad B - V = 1.13 \pm 0.01,$$
$$U - B = 1.18 \pm 0.05.$$

If the averaging is performed not over five nights but over 19 individual estimates, then the same mean values of *V*, *U - B,* and *B - V* are obtained. We see that the error in *B - V* (0.01) is smaller than the error in *V* (0.03). The *B* and *V* magnitudes may change synchronously, which may suggest either the existence of variability or the residual errors in reducing the photometric observations. Note that the derived $B - V$ color index closely coincides with the above value of $B - V = 1.13$ from the literature (Corben 1966).

DETERMINING THE PARAMETERS
$T_{eff}$ AND log $g$

To determine the effective temperature $T_{eff}$ by the photometric method (Lyubimkov and Poklad 2014), we use the derived $U - B$ and $B - V$ and find $Q = 0.37 \pm 0.06$. After substituting this value of $Q$ into Eqs. (1) and (2) from Lyubimkov and Poklad (2014), we obtain two different temperatures $T_{eff}$ corresponding to the metallicities [Fe/H] = 0.0 and -0.5. As can be seen from Table 2, the temperatures $T_{eff}$ = 4430 ± 90 and 4270 + 50 K correspond to the cases of normal, [Fe/H] = 0.0, and reduced, [Fe/H] = -0.5, metallicity, respectively. The discrepancy in $T_{eff}$ is 160 K, and it can affect noticeably the lithium abundance being determined. Thus, to apply the photometric method to obtain a reliable value of $T_{eff}$, it is first necessary to estimate the stellar metallicity.

To determine the effective temperature of HD 77361, we also applied the other (spectroscopic) method (Kovtyukh 2007). As has been noted above, the ratios of the central intensities for pairs of lines with significantly differing excitation potentials are considered in this method. The specific pairs of lines were chosen by Kovtyukh depending on the $T_{eff}$ range. The star HD 77361 falls within the range $T_{eff}$ = 3770-5800 K. Kovtyukh gives six pairs of lines for it, all lines in the spectral range 6040-6155 Å. The spectrum for the cool star HD 77361 is rather blended; nevertheless, three of the six pairs of lines are well suited for applying this method: Ti I 6126.21/Si I 6155.14, V I 6135.36/Si I 6142.49, and V I 6150.16/Si I 6237.32. On average, we obtained the effective temperature $T_{eff}$ = 4310 ± 60 K from these three pairs. Comparing this value with that found by the photometric method (Table 2), we see that it lies between the two values of $T_{eff}$ that this method gives at [Fe/H] = 0 and -0.5.

For each value of $T_{eff}$, Table 2 gives the corresponding surface gravity of the star log $g$. As has been noted above, we determined log $g$ from the stellar parallax. The parallax of HD 77361 is known with a high accuracy: $\pi$ = 9.25 ± 0+43 (van Leeuwen 2007). As a result, the error in log $g$ arising from the error in the parallax is only ±0.02 dex. The total error in log $g$ found by taking into account the uncertainty in $T_{eff}$ is ±0.07 dex at $T_{eff}$ = 4270 ± 50 K and ±0.13 dex at $T_{eff}$ = 4430 ± 90 K.

While applying this method of determining log $g$, we used evolutionary tracks (Claret 2004); thus, we simultaneously determined the stellar mass $M$ (for a description of the technique for determining $M$, see Lyubimkov et al. 2010). The values of $M/M_\odot$ are also given in Table 2.

STELLAR METALLICITY

As has been noted above, the stellar metallicity should be known to reliably determine the effective temperature $T_{eff}$ from the photometric $Q$ index. The metallicity is usually estimated from the iron abundance. In this case, the metallicity index [Fe/H] is found from the equality [Fe/H] = log ε(Fe) - log ε$_\odot$(Fe), where log ε(Fe) and log ε$_\odot$(Fe) are the iron abundances in the stellar and solar atmospheres, respectively. The iron abundance here (just as the lithium abundance above) is given in a standard logarithmic scale, where log ε(H) = 12.00 is adopted for hydrogen. As the solar Fe abundance, we used log ε(Fe) = 7.50 (Asplund et al. 2009).

Just as in LLKPPR'12, we relied on the model atmospheres computed with the SAM12 code (Pavlenko 2003) in our calculations of iron and lithium lines. In accordance with Table 2, we computed two models with the effective temperatures $T_{eff}$ = 4430 and 4270 K that were found by the photometric method at [Fe/H] = 0 and -0.5, respectively.

The iron abundance log ε(Fe) for each of these models was determined from both Fe I and Fe II lines. We used the list of such lines from the paper by Ramirez and Allende Prieto (2011) on Arcturus; the atomic data, including the line excitation potentials and oscillator strengths, were also taken from there. (Arcturus in very similar in its parameters to HD 77361.) The spectral region from 6703 to 6714 Å containing the Li I 6707.8 Å line constitutes an exception; here, the atomic data for all lines, including the Fe I lines, were taken from LLKPPR'12 (see Table 2 in it). Note that the list of lines from Yakovina et al. (2011) forms the basis for these data.

We considered a total of 32 Fe I lines and 8 Fe II lines. The iron abundance log ε(Fe) was determined from the profiles of these lines, i.e., the observed profile of each line was fitted by the synthetic spectrum computed near this line. This method of estimating log ε(Fe) is more reliable than the determination of log ε(Fe) from the line equivalent widths $W$. We used the $W$ measurements only to determine the microturbulence $V_t$; in this case, we applied the standard method consisting in choosing $V_t$ at which there was no trend in log ε(Fe) with increasing $W$. Note that the values of $W$ for the Fe I and Fe II lines in HD 77361 lie within the ranges from 54 to 123 mÅ and from 36 to 106 mÅ, respectively.

The mean iron abundance found from the Fe I lines for both model atmospheres ($T_{eff}$ = 4430 and 4270 K) turned out to be almost the same, log ε(Fe) ≈ 7.40. The Fe abundance found from the Fe II lines also turned out to be almost the same for both models, log ε(Fe) ≈ 7.50. The microturbulence determined from the Fe I and Fe II lines lies within a rather narrow range, $V_t$ = 1.1 -1.3 km s$^{-1}$.

It should be noted that almost the same Fe abundance was obtained for both model atmospheres ($T_{eff}$ = 4430 and 4270 K), despite a noticeable difference (160 K) in their effective temperatures $T_{eff}$. To find an explanation for this phenomenon, we performed the calculations of Fe I and Fe II lines for a number of tabulated model atmospheres from Kurucz (1993) with different parameters $T_{eff}$ and log $g$. Having constructed the dependence of the calculated equivalent widths $W$ on $T_{eff}$ for the Fe I lines, we found that the values of $W$ show a smooth maximum, i.e., the Fe I lines here turned out to be insensitive to slight $T_{eff}$ variations, precisely in the range $T_{eff}$ = 4300-4500 K of interest to us. As regards the Fe II lines, their equivalent widths $W$ in the same $T_{eff}$ range depend very noticeably on $T_{eff}$ (decrease with decreasing $T_{eff}$); at the same time, however, they also show a noticeable dependence on log $g$ (increase with decreasing log $g$). As a consequence, the passage from the model with $T_{eff}$ = 4430 K to the cooler model with $T_{eff}$ = 4270 K, i.e., the reduction in $T_{eff}$ by 160 K, for the Fe II lines is largely compensated for by the simultaneous reduction in log $g$ by 0.23 dex.

The fact that both Fe I and Fe II lines give almost identical iron abundances for HD 77361, irrespective of the model atmosphere under consideration ($T_{eff}$ = 4430 or 4270 K), facilitates considerably our task. Based on the Fe II lines, which are insensitive to departures from LTE, we see that the iron abundance log ε(Fe) ≈ 7.50 found from them coincides with the solar one log ε$_\odot$(Fe) = 7.50 (Asplund et al. 2009). In other words, HD 77361 has a normal metallicity. The Fe I lines showed a slightly reduced Fe abundance (approximately by 0.1 dex), and this can be a consequence of the disregarded departures from LTE (see below).

Thus, we concluded that the star being investigated has a normal metallicity. Therefore, the model with $T_{eff}$ = 4270 K and log $g$ = 2.16 in Table 2, which was obtained under the condition of reduced metallicity [Fe/H] = -0.5, should be excluded from the subsequent consideration. As a result, only two models with the effective temperatures $T_{eff}$ = 4430 ± 90 (found by the photometric method) and 4310 ± 60 K (found by the spectroscopic method) remain in Table 2. The accuracy of both methods for determining $T_{eff}$ is approximately the same, and we cannot give preference to any one of them; therefore, we took the mean $T_{eff}$ = 4370 K as the final one. Having determined other parameters of the star in accordance with this value of $T_{eff}$, we finally obtained the following quantities for HD 77361:

$T_{eff}$ = 4370 ± 100 K,     log $g$ = 2.30 ± 0.10,
log ε(Fe) = 7.49 ± 0.14,     $V_t$ = 1.1 ± 0.2 km s$^{-1}$,
$V$sin$i$ = 4.5 km s$^{-1}$,     $M$ = 1.3 ± 0.2 $M_\odot$.

Here, the iron abundance log ε(Fe) and the microturbulence $V_t$ were found from the Fe II lines, which, in contrast to the Fe I lines, are insensitive to departures from LTE (see below). The projected rotational velocity $V$sin$i$ was determined from the profiles of relatively strong Fe I lines near the Li I 6707.8 Å line. The stellar mass $M$ was estimated using evolutionary tracks (Claret 2004). The stellar age found from the same evolutionary calculations is ~4.5 × 10$^9$ yr with an uncertainty of the order of ±2 × 10$^9$ yr. These errors in the mass and age were obtained by taking into account the errors of ±100 K in $T_{eff}$ and ±0.10 dex in log $g$.

Our effective temperature $T_{eff}$ = 4370 ± 100 K turned out to be lower than $T_{eff}$ = 4580 ± 75 K obtained by Kumar and Reddy (2009) from the Fe I lines by 210 K. The fundamental difference between the methods of determining $T_{eff}$ is responsible for such a discrepancy. Above, we discussed in detail these differences. It should be noted that Kumar and Reddy estimated $T_{eff}$ not only spectroscopically, from the Fe I lines, but also photometrically, from the $B$ - $V$ and $V$ - $K$ color indices (they found $T_{eff}$ = 4550 and 4587 K, respectively). In doing so, they used the calibrations from Alonso et al. (1999), in which the averaged dependence of $T_{eff}$ on $B$ - $V$ or $V$ - $K$ was represented by a quadratic polynomial dependent on two variables, the $B$ - $V$ (or $V$ - $K$) color index itself and the metallicity index [Fe/H]. We repeated their calculation of $T_{eff}$ at $B$ - $V$ = 1.13 and [Fe/H] = -0.09 (corresponding to the Fe I lines); as a result, we obtained $T_{eff}$ = 4543 K. The latter value is higher than the temperature $T_{eff}$ = 4370 ± 100 K we adopted by 173 K.

Alonso et al. (1999) provided the standard deviation $σ(T_{eff})$ = 96 K for their calibration in the case of $B$ - $V$; however, as can be seen from Fig. 2, the scatter of individual stars around the averaged dependence reaches ±200 K. It should be emphasized that such averaged calibrations are constructed from normal stars. However, the phenomenon of super Li-rich stars still remains a puzzle (see below); therefore, the star HD 77361 cannot be considered quite normal. We think that the difference of 173 K between our determination of $T_{eff}$ for HD 77361 and the result obtained from the calibration of Alonso et al. (1999) could be explained in principle by three factors: (1) the inaccuracies of this calibration for individual stars; (2) the possible mismatch between the "normal" star HD 77361 and the average dependence for normal stars; and (3) the error of our $T_{eff}$ determination. However, as we will show below, some uncertainty in $T_{eff}$ does not affect in any way our final conclusion that the giant HD 77361 belongs to the type of super Li- rich stars.

DEPARTURES FROM LTE IN IRON LINES

While discussing the metallicity of HD 77361, the possible non-LTE effects in the iron lines for this star should be considered at least briefly. We found the iron abundance from 32 Fe I lines and 8 Fe II lines based on the model atmosphere computed for $T_{eff}$ = 4370 K and log $g$ = 2.30 to be log ε(Fe) = 7.41 ± 0.10 and 7.49 ± 0.14, respectively. The microturbulence $V_t$ = 1.3 ± 0.2 km s$^{-1}$ for Fe I and $V_t$ = 1.1 ± 0.2 km s$^{-1}$ for Fe II corresponds to these values of log ε(Fe). For comparison, note that Pakhomov et al. (2009) found $V_t$ = 1.19-1.52 km s$^{-1}$ with a typical error of 0.15 km s$^{-1}$ for 26 G and K giants from the iron lines in the LTE approximation.

The difference in the abundances log ε(Fe) that we found for HD 77361 from the Fe I and Fe II lines is -0.08 dex. On the one hand, the difference is insignificant and comparable to the error in log ε(Fe). On the other hand, it can be real and be explained by the disregarded departures from LTE in the Fe I lines. This is suggested by some data from the literature.

As early as 30 years ago, Boyarchuk et al. (1985) found that the Fe I lines in F-type supergiants are subjected to significant departures from LTE. This was shown to be explained by significant overionization (relative to LTE) of Fe I atoms in the atmospheres of such stars due to the ultraviolet radiation coming from below. It was established that neglecting the non-LTE effects leads to a noticeable underestimation of the iron abundance determined from the Fe I lines: up to 0.5 dex for relatively strong Fe I lines in the case of F0 supergiants and up to 0.2 dex in the case of cooler F8 supergiants (Lyubimkov and Boyarchuk 1983). Hence it ostensibly followed that the role of non-LTE effects in the Fe I lines decreased with decreasing effective temperature $T_{eff}$. As regards the Fe II lines, on the contrary, they turned out to be insensitive to non-LTE effects. The following question arises: Are so significant non-LTE effects in the Fe I lines possible for cooler K giants and, in particular, for HD 77361?

Subsequent non-LTE calculations for cooler stars confirmed the main conclusions of Boyarchuk et al. (1985) concerning primarily the Fe overionization in the atmospheres of such stars and the insensitivity of Fe II lines to departures from LTE. The calculations for 136 G- and K-type stars in a wide range of [Fe/H] (these were predominantly dwarfs) that were performed by Thevenin and Idiart (1999) can be cited as an example. The non-LTE corrections to the iron abundance found from the Fe I lines were shown in this paper to exhibit a clear dependence on the metallicity index [Fe/H]: they increase with decreasing [Fe/H]. In particular, this correction is ~0.3 dex at [Fe/H] = -3.0 dex, while it is close to zero at normal metallicity [Fe/H] = 0. It was confirmed that the Fe II lines are not subjected to noticeable non-LTE effects.

The paper by Ramirez et al. (2013), where a large group (more than 800) of F, G, and K stars was investigated, is also of interest; just as in the previous paper, these were predominantly dwarfs, but mostly with a normal metallicity. The difference in iron abundance between the Fe II and Fe I lines, which, as follows from the above discussion, is actually the non-LTE correction for the Fe I lines, was determined for these stars. It turned out to be close to zero, on average, for all stars; however, it was, on average, 0.09 ± 0.07 dex for cool K stars with temperatures $T_{eff}$ < 4800 K (the $T_{eff}$ range of greatest interest to us).

In the case of the K giant HD 77361 being investigated, the non-LTE calculations of Fe I and Fe II lines performed by Takeda (1991) for Arcturus are of particular interest; this is also an early K giant with a similar effective temperature, $T_{eff}$ = 4286 K (Ramirez and Allende Prieto 2011). By varying some parameters in the non-LTE calculations, Takeda found that, depending on these variations, the mean non-LTE correction for the Fe I lines changes from 0 to 0.1 dex, while the individual corrections for these lines are always less than 0.2 dex. It should be noted that there can be negative non-LTE corrections occasionally reaching -0.1 dex for the Fe II lines. Regarding Arcturus, it should also be noted that by investigating the observed spectrum of this star, Rami rez and Allende Prieto (2011) obtained a difference of 0.12 dex in Fe abundance between the Fe II and Fe I lines, which is comparable to our difference of 0.08 dex for HD 77361.

All of the above data from the literature, including the results for Arcturus, lead us to conclude that the inferred difference of -0.08 dex in iron abundance between the Fe I and Fe II lines for the K giant HD 77361 can be entirely attributed to the disregarded non-LTE corrections for the Fe I lines.

DETERMINING THE LITHIUM ABUNDANCE

When determining the lithium abundance, one often restricts oneself to considering the strongest, resonance Li I 6707.8 Å line, because the weaker Li I lines are usually unseen. Apart from the strong 6707.8 Å line, the subordinate Li I 6103.6 Å line is clearly seen in the spectrum of HD 77361 owing to its low temperature and high lithium abundance; the Li I 8126.4 Å line is also seen in the blend. We also used these two lines to estimate the lithium abundance.

Each of the three Li I lines considered is actually a blend consisting of several components, and this fact was taken into account in our calculations. The parameters of these components are given in Table 3, including the wavelength A, lower-level excitation potential $E_l$, and oscillator strength log $gf$. Note that these data refer to the lithium isotope $^7$Li; we neglected the contribution from the much less abundant isotope $^6$Li (recall that $^6$Li/$^7$Li = 0.08 for the Sun; Asplund et al. 2009).

Table 3. Data for the investigated Li I lines and the lithium abundance found from them

| Li I blend | Blend components | | | log $\varepsilon$(Li) | log $\varepsilon$(Li) |
|---|---|---|---|---|---|
| $\lambda$, Å | $\lambda$, Å | $E_l$, eV | log $gf$ | LTE | non-LTE |
| 6707.8 | 6707.754 | 0.000 | −0.431 | 3.80 | 3.69 |
|  | 6707.766 | 0.000 | −0.209 |  |  |
|  | 6707.904 | 0.000 | −0.733 |  |  |
|  | 6707.917 | 0.000 | −0.510 |  |  |
| 6103.6 | 6103.538 | 1.848 | 0.101 | 3.38 | 3.75 |
|  | 6103.649 | 1.848 | 0.361 |  |  |
|  | 6103.664 | 1.848 | −0.599 |  |  |
| 8126.4 | 8126.231 | 1.848 | −0.665 | 3.60 | 3.80 |
|  | 8126.452 | 1.848 | −0.365 |  |  |

The lithium abundance was determined by computing the synthetic spectra and fitting the computed profile of each Li I blend to the observed spectrum. The technique of our non-LTE calculations is described in LLKPPR'12. In particular, We computed a model atmosphere corresponding to the adopted parameters $T_{eff}$, log $g$, log $\varepsilon$(Fe), and $V_t$ using the SAM12 code (Pavlenko 2003). It should be noted that to calculate the strong resonance Li I 6707.8 Å line, we had to extrapolate the computed model atmosphere by extending it to high layers (see below).

In our computations of the synthetic spectra, we used the microturbulence $V_t = 1.1$ km s$^{-1}$ found above from the Fe II lines and the projected rotational velocity $V\sin i = 4.5$ km s$^{-1}$ determined from relatively strong Fe I lines near the Li I 6707.8 Å line. The observed and computed spectra near the three Li I lines are compared in Figs. 1-3. Here, the thick solid curve indicates the observed spectrum, while the dashed curve indicates the synthetic spectrum convolved with rotation at $V\sin i = 4.5$ km s$^{-1}$. As has been noted above, when computing the synthetic spectrum near Li I 6707.8 Å, we used the data from LLKPPR'12 for the lines. For the lines near Li I 6103.6 and 8126.4 Å, we took the data from the VALD3 database (Ryabchikova et al. 2015).

It can be seen from Fig. 1 that the star's spectrum consists of fairly sharp and narrow lines due to its low rotational velocity. Apart from the strong Li I 6707.8 Å blend, five more relatively strong Fe I (marked in Fig. 1) are present in Fig. 1. When calculating these five lines, we used the same iron abundance $\log \varepsilon(\mathrm{Fe}) = 7.41$; it is this value that was found above from 32 Fe I lines. Several weak CN molecular lines can also be seen in Fig. 1 to the right of the Li I 6707.8 Å line. The synthetic spectrum poorly reproduces these features; there are shifts in wavelength that can be explained by the inaccuracy of the available data for such weak CN lines. Note that in our calculations of these CN lines, we adopted the C and N abundances obtained below by analyzing the CN + Li I blend at 8126.4 Å (see the comment to Fig. 3).

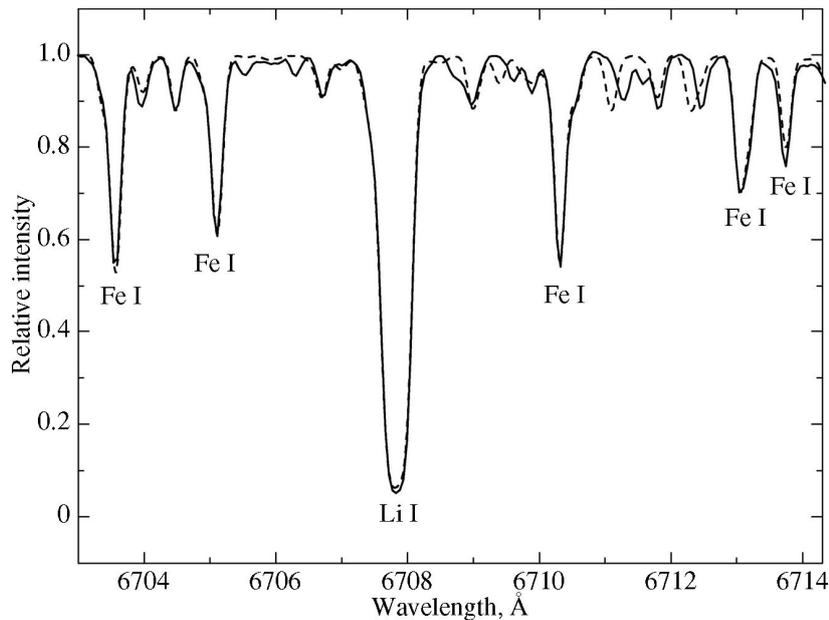

**Fig. 1.** Comparison of the observed (thick solid curve) and synthetic (dashed curve) spectra near the Li I 6707.8 Å line. In addition to the Li I line, the positions of five more Fe I lines are indicated.

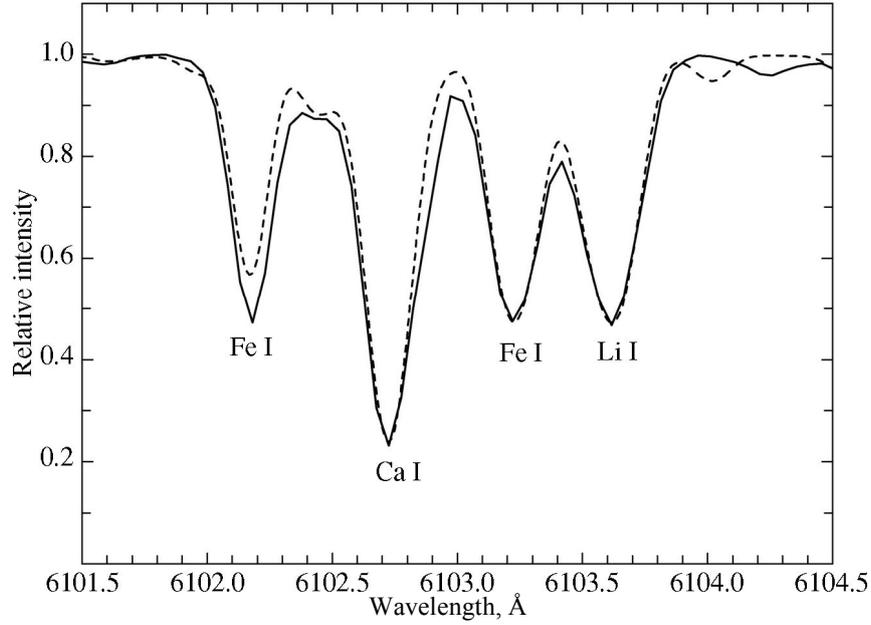

**Fig. 2.** Comparison of the observed (thick solid curve) and synthetic (dashed curve) spectra near the Li I 6103.6 Å line. In addition to the Li I line, the positions of two more Fe I lines and one Ca I lines are indicated.

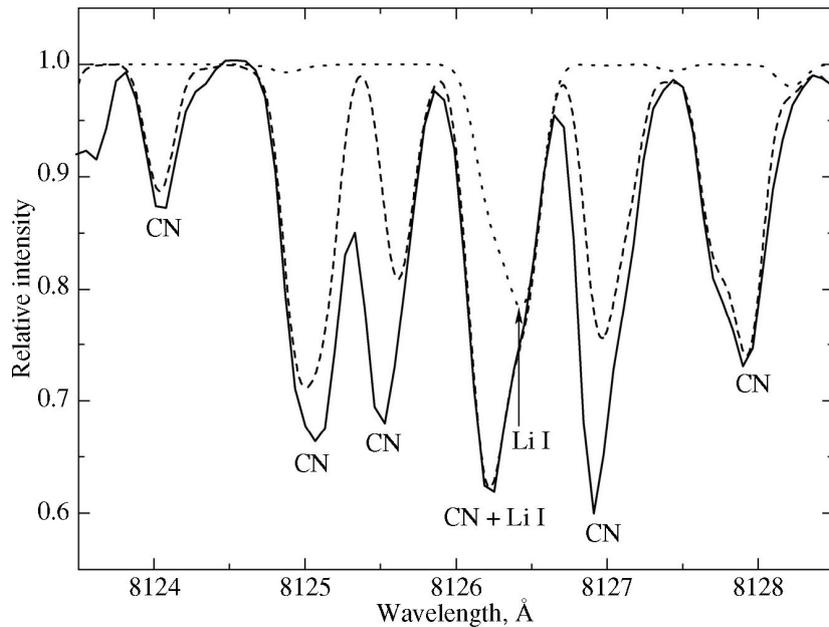

**Fig. 3.** Comparison of the observed (thick solid curve) and synthetic (dashed curve) spectra near the Li I 8126.4 Å line. In addition to the CN + Li I blend, the positions of five more CN blends are indicated. The dotted curve indicates the synthetic spectrum computed by taking into account only the atomic lines.

It can be seen from Fig. 1 that the resonance Li I 6707.8 Å line is deep and strong; its observed central depth is 95%, and its equivalent width $W = 520$ mÅ. The core of this line is formed in the high rarefied layers of the stellar atmosphere; as has been noted above, we had to extrapolate the initial model atmosphere to calculate the line core. We extrapolated the model to the high atmospheric layers by the method proposed by Pavlenko (1984); more specifically, we assumed that these layers were described by a polytropic equation and that the polytropic index could be calculated from several upper layers of the computed model atmosphere. As can be seen from Fig. 1, a slight discrepancy of ~1% between the computed and observed profiles, nevertheless, remained at the line center after such an extrapolation, although there is complete agreement in the line wings.

Table 4. Comparison of our lithium abundances log ε(Li) (LTE and non-LTE) and non-LTE corrections Δ with the results of Kumar and Reddy (2009)

| Li I line | Kumar and Reddy (2009) | | | Our values | | |
|---|---|---|---|---|---|---|
| λ, Å | LTE | Δ | non-LTE | LTE | Δ | non-LTE |
| 6707.8 | 3.96 | −0.16 | 3.80 | 3.80 | −0.11 | 3.69 |
| 6103.6 | 3.67 | +0.16 | 3.83 | 3.38 | +0.37 | 3.75 |

Our approximate extrapolation is probably insufficient when modeling so high atmospheric layers. The approach where a spherical model is considered instead of the standard plane-parallel atmosphere seems more realistic here. In this case, the sphericity effect, which may turn out to be significant for the strong 6707.8 Å line, must be negligible for weaker lines (W < 300 mÅ) in HD 77361. This follows from the calculations of Heiter and Eriksson (2006), according to which the extent of the atmosphere for a K giant with such parameters $T_{eff}$, log g, and $M$ in the line formation region (between the optical depths $\tau_R = 1$ and $10^{-5}$) is only ~1% of the stellar radius.

Three Fe I and Ca I lines (marked in Fig. 2) are seen in Fig. 2, where the Li I 6103.6 Å doublet region is considered, to the left of this Li I line. In our calculations of the two Fe I lines, we adopted, as above, the iron abundance log ε(Fe) = 7.41. As regards the Ca I 6102.72 line, it is even stronger than the Li I 6103.6 Å line being investigated (the equivalent widths of these two lines are $W$ = 204 and 138 mÅ, respectively). For this Ca I line, we achieved the best agreement between the observed and computed profiles at the calcium abundance log ε(Ca) = 5.86. The latter value is lower than the solar abundance log ε(Ca) = 6.34 (Asplund et al. 2009) approximately by 0.5 dex, which, at first glance, seems strange for a star with a normal metallicity. We suggest that such an underestimate of log ε(Ca) is explained by the disregarded departures from LTE.

The non-LTE calculations of Ca I lines performed by Drake (1991) for cool stars can be mentioned in this connection. In particular, for the model atmosphere with $T_{eff}$ = 4500 K, log $g$ = 2.0, and [Fe/H] = 0, which are close to the parameters of our star, he obtained non-LTE corrections of 0.2–0.3 dex to the Ca abundance for the Ca I lines belonging to multiplets 18, 19, and higher. These are lines with lower-level excitation potentials $E_l$ > 2.5 eV. Since the Ca I 6102.72 line under consideration (multiplet 3) has a lower excitation potential, $E_l$ = 1.88 eV, one might expect a more significant non-LTE correction to the calcium abundance for it. It should be noted that the problem of the Ca I 6102.72 line does not affect in any way the lithium abundance determination from the Li I 6103.6 Å line.

The lithium abundances (both LTE and non-LTE) found from the Li I 6707.8 and 6103.6 Å lines are given in Table 3. We see that the non-LTE abundances (3.69 and 3.75, respectively) show excellent agreement. It should be noted that these two lines were considered by Shavrina et al. (2006) when investigating five magnetic roAp stars (these are dwarfs with effective temperatures $T_{eff}$ = 6600–7750 K). Here, the Li abundance from the 6103.6 line was found to be systematically higher (by 0.2–0.4 dex) than that from the 6707.8 line. In the case of our cooler giant HD 77361, there is no significant difference in Li abundance between the 6707.8 and 6103.6 lines.

Figure 3 shows the near-infrared spectral region containing the Li I 8126.4 Å line (to be more precise, the Li I 8126.45 + .23 Å doublet). Here, the situation differs radically from that in Figs. 1 and 2: (1) the Li I 8126.4 Å doublet lies in the right wing of a fairly strong CN molecular line and is not a separate feature in the spectrum; (2) there are no strong atomic lines (except the Li I 8126.4 Å line itself) in this region, and all six observed blends belong to CN. The synthetic spectrum consisting only of atomic lines and convolved with rotation at $V\sin i$ = 4.5 km s$^{-1}$ is indicated in Fig. 3 by the dotted curve. Note that the total equivalent width of the CN + Li I blend in Fig. 3 is 165 mÅ, while the equivalent width of the Li I 8126.4 Å line itself (indicated by the arrow) is 85 mÅ, i.e., its contribution to the full blend is 52%.

A careful calculation of the CN line at 8126.2 Å is particularly important in our case. When fitting the synthetic spectrum to the observed CN + Li I blend (to be more precise, to its left wing), we took the carbon and nitrogen abundances to be $\log \varepsilon(C) = 8.42$ and $\log \varepsilon(N) = 8.37$, respectively. This C abundance was determined from two C2 lines and one C I line; it turned out to be close to the solar value of $8.50 \pm 0.06$ (Caffau et al. 2010). The N abundance was derived directly from the CN 8126.2 Å line; it should be considered as a preliminary estimate. As can be seen from Fig. 3, the observed CN blends at 8124 and 8128 Å are satisfactorily described at the same C and N abundances; however, the CN blends at 8125, 8125.5, and 8127 Å turned out to be noticeably deeper than the computed ones. Apart from the errors in the adopted C and N abundances, insufficient completeness and accuracy of the data for the CN lines in the spectral region under consideration can be responsible for such a discrepancy. If, following other authors, the CN lines near 8000 Å are used to estimate the N abundance, then we obtain $\log \varepsilon(N) = 7.96$ close to the solar abundance of $7.86 \pm 0.12$ (Caffau et al. 2009). A more detailed analysis of the C and N abundances for HD 77361 is a separate problem that is beyond the scope of this paper.

Since the Li I 8126.4 Å line is observed in a blend, one might expect the lithium abundance to be determined from it not so reliably as from the 6707.8 and 6103.6 Å lines. Nevertheless, the non-LTE abundance $\log \varepsilon(Li) = 3.80$ derived from the Li I 8126.4 Å line turned out to be very close to $\log \varepsilon(Li)$ found from the two previous lines (Table 3).

It is interesting to compare the LTE and non-LTE values of $\log \varepsilon(Li)$. According to Table 3, the difference between the non-LTE and LTE cases is -0.11 dex for the resonance line at 6707.8 Å and +0.37 and +0.20 dex, respectively, for the subordinate lines at 6103.6 and 8126.4 Å. We see that the non-LTE corrections to the lithium abundance for the K giant under consideration are fairly large. As has already been noted, the non-LTE corrections depend on the parameters of a specific star; in particular, they depend on $T_{eff}$ and can reach 0.4 dex for the Li I 6707.8 Å line (LLKPPR'12).

Based on $\log \varepsilon(Li)$ found for the three Li I lines considered (Table 3), we obtain the mean non-LTE lithium abundance $\log \varepsilon(Li) = 3.75 \pm 0.05$ for HD 77361. Given the possible uncertainty of ±100 K in the effective temperature, which makes the dominant contribution to the total error in $\log \varepsilon(Li)$, we finally obtain $\log \varepsilon(Li) = 3.75 \pm 0.11$.

## DISCUSSION

The derived lithium abundance $\log \varepsilon(Li) = 3.75 \pm 0.11$ exceeds considerably the initial abundance $\log \varepsilon(Li) = 3.2 \pm 0.1$ for young stars in the solar neighborhood. Thus, we confirmed that the K giant HD 77361 actually belongs to the type of super Li-rich stars.

As has been noted in the Introduction, Kumar and Reddy (2009, hereafter KR'09) found the mean lithium abundance $\log \varepsilon(Li) = 3.82$ for this star from the Li I 6707.8 and 6103.6 Å lines. Our mean non-LTE abundance $\log \varepsilon(Li) = 3.75$ is smaller than their value only by 0.07 dex, but our analysis of the Li abundances from individual lines revealed more significant differences. This can be seen from Table 4, where the lithium abundances (both LTE and non-LTE) derived from the Li I 6707.8 and 6103.6 Å lines in KR'09 and our work are compared.

The effective temperature $T_{eff} = 4580$ K adopted in KR'09 for HD 77361 is higher than our value by 210 K. As our calculations showed, a rise in $T_{eff}$ by 210 K (other things being equal) must cause the LI abundance to increase by 0.26 dex for the 6707.8 line and by 0.23 dex for the 6103.6 line. As can be seen from Table 4, the actual changes differ from these values: the increase in LTE Li abundance in KR'09 is 0.16 dex for the 6707.8 line and 0.29 dex for the 6103.6 line. Therefore, other features that distinguish our study of HD 77361 from KR'09 are also noteworthy.

In particular, the model atmospheres computed with R. Kurucz's 1994 code (http://kurucz.harvard.edu/) were used in KR'09, while we applied the more perfect model atmospheres that were computed by one of us using our own code (Pavlenko 2003) developed specially to investigate cool stars. Next, KR'09 estimated the non-LTE corrections Δ to the lithium abundance for the Li I 6707.8 and 6103.6 Å lines using published data, while we applied our own, repeatedly tested technique of non-LTE calculations of Li I lines. A slight reduction in the Li abundance (~0.1 dex for the 6707.8 line) could take place in KR'09 due to the microturbulence V being overestimated to 1.4 km s$^{-1}$ compared to our value 1.1 km s$^{-1}$. It can be concluded that the difference in $\log \varepsilon(Li)$ between our work and KR'09 is explained by a combination of several factors; in other words, it is a consequence of the differences in investigation technique.

Thus, using the investigated K giant HD 77361 as an example, we confirmed the real existence of rare super Li-rich stars. The red giant no. 3416 in the open cluster Trumpler 5, which, according to Monaco et al. (2014), also belongs to the group of super Li-rich stars and has a lithium abundance $\log \varepsilon(Li) = 3.75 \pm 0.10$ coincident with the abundance that we found for HD 77361, can serve as another confirmation of this fact. Note that $\log \varepsilon(Li) = 3.75$ in the above paper was derived from a non-LTE analysis of the Li I 6707.8 and 6103.6 Å lines. Interestingly, an enhanced lithium abundance in this star (its parameters are $T_{eff} = 5000$ K and $\log g = 2.5$) is accompanied by a low ratio of the carbon isotopes, $^{12}C/^{13}C = 14$ (for the Sun, $^{12}C/^{13}C = 89$; Asplund et al. 2009). Thus, our giant HD 77361 is also similar to this star in that a low ratio, $^{12}C/^{13}C = 4$ (KR'09), was also obtained for HD 77361. The greatly reduced values of $^{12}C/^{13}C$ confirm that both giants have passed the phase of deep convective mixing.

Under such mixing, all of the lithium contained in the atmosphere must have been depleted. The fact that, on the contrary, an unusually high lithium abundance is observed here suggests that this lithium has recently been synthesized.

The assumption about a recent Li synthesis has to be invoked to also explain the enhanced Li abundance in Li-rich stars, which do not have such a strong excess of this element; log ε(Li) in them varies approximately from 2.0 to 3.3, i.e., it does not show an appreciable excess relative to the initial value of log ε(Li) = 3.2 ± 0.1. It should be emphasized that the standard theory does explain the observed enhanced values of log ε(Li) for some of such stars; these are giants with a low initial rotational velocity, $V_0 \sim 0$ km s$^{-1}$, that have not yet reached the phase of deep mixing. Examples of two such stars are given in LLKPPR'12: the F giants HR 7008 and HD 17905 that have retained the initial lithium abundance log ε(Li) = 3.2 ± 0.1.

Other Li-rich stars in which the observed Li excess cannot be explained by theoretical models are also pointed out in LLKPPR'12. In particular, in some giants with masses $M < 6 M_\odot$, an enhanced lithium abundance is accompanied either by a significant rotational velocity or by a nitrogen excess; according to the standard theory, these features are incompatible with the high values of log ε(Li). Apart from a nitrogen excess, the giant HR 3102 exhibits a carbon deficiency (Lyubumkov et al. 2015); these anomalies in the C and N abundances suggest once again that deep convective mixing has occurred in the star (by the anomalies here we mean the deviations from the solar (initial) C and N abundances). The observed high Li abundance in some of the Li-rich stars in combination with the anomalous C and N abundances and the low $^{12}C/^{13}C$ ratio (Kumar et al. 2011) can be explained only by a recent synthesis of lithium in them.

The Cameron-Fowler mechanism (Cameron and Fowler 1971) is invoked to explain the synthesis of lithium in red giants. It includes the following reactions: $^3He + \alpha \to {}^7Be + \gamma$; $^7Be + e^- \to {}^7Li + \nu_e$, i.e., $^7Li$ atoms are synthesized from $^3He$ via $^7Be$. The Cameron—Fowler mechanism takes place under rather peculiar conditions; in particular, convection, which contributes to the rapid dredge-up of $^7Be$ into cooler atmospheric layers, must play an important role. Initially, this mechanism was proposed to explain the synthesis of lithium in AGB (Asymptotic Giant Branch) stars with masses $M \approx 4-6 M_\odot$; subsequently, it was applied to explain the lithium enrichment of the Galaxy through nova eruptions (Romano et al. 2001; D'Antona and Ventura 2010). As regards the Li-rich RGB (Red Giant Branch) stars with masses $M \approx 1-2 M_\odot$, to which the star being investigated also belongs, the application of the Cameron-Fowler mechanism here runs into certain difficulties. Our task does not include the discussion of these difficulties; it can only be noted that in the opinion of some authors, the phenomenon of Li-rich K giants still remains a puzzle (Uttenthaler et al. 2012).

Only the fact that the phase of lithium synthesis in FGK giants and supergiants is very short in duration is beyond doubt; this is evidenced by the very small fraction (~1%) of Li-rich and super Li-rich stars among them. This fact should be taken into account when constructing the hypotheses that explain the phenomenon of Li-rich and super Li-rich stars.

CONCLUSIONS

Let us summarize the results of our study of the K giant HD 77361.

To determine its effective temperature $T_{eff}$, we used two independent methods, photometric (Lyubimkov and Poklad 2014) and spectroscopic (Kovtyukh 2007). To apply the photometric method, we performed $UBV$ observations of the star, which yielded $V = 6.18 \pm 0.03$, $B - V = 1.13 \pm 0.01$, and $U - B = 1.18 \pm 0.05$.

Applying the photometric method required a preliminary study of the metallicity in HD 77361; we established that the star has a normal (solar) iron abundance. While studying the Fe abundance, we showed that the departures from LTE in the Fe I lines for this K giant are small and correspond to a non-LTE correction of ~0.1 dex in the Fe abundance.

The accuracy of both methods for determining $T_{eff}$ is approximately the same, and we could not give preference to any one of them; therefore, the mean value of $T_{eff} = 4370 \pm 100$ K was taken as the final one. This value is lower than the effective temperature $T_{eff}$ that was found previously for HD 77361 by Kumar and Reddy (2009) by 210 K.

The following parameters of the star correspond to the effective temperature $T_{eff}$ found: surface gravity log $g$ = 2.30 ± 0.10, iron abundance log ε(Fe) = 7.49 ± 0.14, microturbulence $V_t$ = 1.1 ± 0.2 km s$^{-1}$, rotational velocity $V\sin i$ = 4.5 km s$^{-1}$, and mass $M$ = 1.3 ± 0.2 $M_\odot$. The age of the star is ~4.5 × 10$^9$ yr with an uncertainty of the order of ±2 × 10$^9$ yr.

We determined the lithium abundance from a non-LTE analysis of three Li I lines: the resonance line at 6707.8 Å and the subordinate lines at 6103.6 and 8126.4 Å; the latter is in a blend with a CN molecular line. We obtained the lithium abundance log ε(Li) = 3.75 ± 0.11, which exceeds considerably the initial abundance log ε(Li) = 3.2 ± 0.1 for young stars in the solar neighborhood. Thus, based on our own independent technique, we confirmed the conclusion reached by Kumar and Reddy that the K giant HD 77361 belongs to the type of super Li-rich stars.

The observed high lithium abundance in cool super Li-rich giants and supergiants is inconsistent with the standard stellar evolution theory. It has to be assumed that there has been a recent synthesis of lithium in such giants (including HD 77361). This episode in the evolution of such stars was probably short in duration. This problem requires a further study.

ACKNOWLEDGMENTS

We are grateful to D.L. Lambert for the spectrum of the giant HD 77361 placed at our disposal.